\documentclass[aps,twocolumn,showpacs]{revtex4}
\usepackage[dvips]{graphicx}
\newcommand{\sect}[1]{\setcounter{equation}{0}\section{#1}}

\begin{document}
\title{Power-Law tailed statistical distributions and Lorentz transformations}
\author{G. Kaniadakis}
\email{giorgio.kaniadakis@polito.it}
\affiliation{Dipartimento di Fisica, Politecnico di Torino, \\
Corso Duca degli Abruzzi 24, 10129 Torino, Italy}
\date{\today}

\begin {abstract}

The present Letter, deals with the statistical theory [Phys. Rev. E {\bf 66}, 056125 (2002) and Phys. Rev E {\bf 72}, 036108 (2005)], which predicts the probability distribution $p(E) \propto \exp_{\kappa} (-I)$, where, $I \propto \beta E -\beta \mu$, is the collision invariant, and  $\exp_{\kappa}(x)=(\sqrt{1+ \kappa^2 x^2}+\kappa x)^{1/\kappa}$, with $\kappa^2<1$. This, experimentally observed distribution, at low energies behaves as the Maxwell-Boltzmann exponential distribution, while at high energies presents power law tails. Here we show that the function $\exp_{\kappa}(x)$  and its inverse $\ln_{\kappa}(x)$, can be obtained within the one-particle relativistic dynamics, in a very simple and transparent way,  without invoking any extra principle or assumption, starting directly from the Lorentz transformations. The achievements support the idea that the power law tailed distributions are enforced by the Lorentz relativistic microscopic dynamics, like in the case of the exponential distribution which follows from the Newton classical microscopic dynamics.

\end {abstract}


\maketitle

\sect{Introduction}

The origin of the power-law tails in the particle
population, observed in a wide variety of systems from high energy physics
to condensed matter physics, is one of
the most pressing outstanding issues in statistical physics. For
instance the cosmic ray spectrum, which extends over 13 decades in
energy, from $10^8$ eV to $10^{20}$ eV, and spans 33 decades in
particle flux, from $10^{-29}$ to $10^4$ units, is not
exponential at relativistic energies \cite{PRE02,PRE05,EPL10}. More precisely, the particle
spectrum $f_i\propto \chi(\beta E_i-\beta \mu)$ obeys the Boltzmann
law of classical statistical mechanics i.e. $\chi(x)
{\atop\stackrel{\textstyle\sim}{\scriptstyle x\rightarrow \,0}}
\exp(-x)$ for low energies, while for high energies this spectrum
presents power law fat tails i.e $\chi(x)
{\atop\stackrel{\textstyle\sim}{\scriptstyle x\rightarrow
\,+\infty}} x^{-a}$,  the spectral index being close to 2.7-3.1.

Also theoretical problems present
difficulties when treated within the classical statistical
mechanics. For instance the Bekenstein-Hawking area law problem,
related to the black hole entropy, asserts that the entropy of a
black hole scales as the area of the event horizon
\cite{PRE05}. It is well known that the ordinary Boltzmann
statistical mechanics and thermodynamics fail to justify this law,
and conversely predict that the entropy scales as the volume of the
spatial region delimited by the event horizon.

In experimental physics, the cosmic rays problem  was
approached for the first time in 1968, by using a different
distribution from the Boltzmann one. In his proposal Vasyliunas
heuristically identified the function $\chi(x)$ with the Student
distribution function which presents power-law tails
\cite{EPJB2009}. In the last 40 years  a vast amount of
literature has been produced, based on the Vasyliunas distribution,
including both experimental and theoretical works regarding statistical
systems violating the Boltzmann statistics.

On the other hand, in the last few decades, the power-tailed statistical distributions have been observed in a variety of physical, natural or artificial systems. In \cite{EPJB2009} one can find updated discussion of power-law tailed distributions and an extensive list of systems where that distributions have been empirically observed.

In several papers specific physical mechanisms have been explored in
order to furnish theoretical support to the experimentally observed
power-law-tailed distribution functions. However there is
currently an intense debate regarding the theoretical foundations of
the non-Boltzmannian distributions, especially in physics. In the
last  years, after noting that the power-law tails are placed in the high energy region, and then regards relativistic particles, the question has been posed whether the solution of the problem, i.e. the theoretic determination of the function $\chi(x)$ and consequently of the related distribution and
entropy, can be explained by invoking the basic principles of
special relativity.

The present Letter deals with the statistical theory, predicting for the function $\chi(x)$ the very simple form i.e. $\chi(x)=\exp_{\kappa}(x)$ with
\begin{equation}
\exp_{\kappa}(x)=(\sqrt{1+ \kappa^2 x^2}+\kappa x)^{1/\kappa}.
\label{A1}
\end{equation}
The parameter $\kappa^2<1$ is the reciprocal of light speed, in a dimensionless form, while $1/\kappa^2$ is the particle rest energy. In the classical limit $\kappa \rightarrow 0$, $\exp_{\kappa}(x)$ reduces to the ordinary exponential function, defining the Boltzmann factor of classical statistical mechanics \cite{PhA01,PLA01,EPJA2009}.

Starting from the inverse function of $\exp_{\kappa}(x)$, namely $\ln_{\kappa}(x)$ given by
\begin{eqnarray}
\ln_{\kappa}(x)= \frac{x^{\kappa}-x^{-\kappa}}{2\kappa} \ ,
\label{A2}
\end{eqnarray}
the entropy of the system assumes the form
\begin{eqnarray}
S=- \int d^3p \,\,\, f \,\ln_{\kappa}(f) \ .
\label{A3}
\end{eqnarray}
The above entropy, after maximization under the constraints expressing the conservation of the distribution norm and of the system energy, yields the following expression for the distribution function
\begin{eqnarray}
f=\frac{1}{\epsilon}\,\exp_{\kappa}\left(- \gamma \beta\,[E-\mu] \right) \ , \ \ \ \ \ \
\ \ \ \label{A4}
\end{eqnarray}
where $E$ is the particle energy, $\beta$ and $\beta \mu$ are the Lagrange multipliers, while
\begin{eqnarray}
\gamma=\frac{1}{\sqrt{1-\kappa^2}} \ ,  \ \ \ \ \ \ \ \epsilon=\exp_{\kappa}(\gamma)  , \ \ \ \ \ \ \ \label{A5}
\end{eqnarray}
are two constants, depending exclusively on the value of the parameter $\kappa$.

In the last few years various authors have considered the
foundations of this statistical theory, e.g. the H-theorem and the
molecular chaos hypothesis \cite{Silva06A,Silva06B}, the
thermodynamic stability \cite{Wada1,Wada2}, the Lesche stability
\cite{KSPA04,AKSJPA04,Naudts1,Naudts2}, the Legendre structure of
the ensued thermodynamics \cite{ScarfoneWada,Yamano}, the geometrical structure \cite{Pistone}, and other mathematical properties of the theory  \cite{OikPOMP10}, etc. On the other hand
specific applications to physical systems have been considered, e.g.
the cosmic rays \cite{PRE02}, relativistic \cite{GuoRelativistic}
and classical \cite{GuoClassic} plasmas  in presence of external
electromagnetic fields, the relaxation in relativistic plasmas under
wave-particle interactions \cite{Lapenta,Lapenta2009}, anomalous diffusion \cite{WadaScarfone2009,Wada2010}, kinetics
of interacting atoms and photons \cite{Rossani}, particle kinetics in the presence of temperature gradients \cite{GuoDuoTgradient},  particle systems in external conservative force fields \cite{Silva2008}, stellar distributions in astrophysics \cite{Carvalho,Carvalho2,CarvalhoEPL10}, quark-gluon
plasma formation \cite{Tewel}, quantum hadrodynamics models
\cite{Pereira}, the fracture propagation \cite{Fracture}, etc. Other applications regard
dynamical systems at the edge of chaos \cite{Corradu,Tonelli,Celikoglu},
fractal systems \cite{Olemskoi,OlemskoiEPJB10}, the random matrix theory
\cite{AbulMagd,AbulMagd2009}, the error theory \cite{WadaSuyari06}, the game
theory \cite{Topsoe}, the information theory \cite{WadaSuyari07},
etc. Also applications to economic systems have been considered e.g.
to study the personal income distribution
\cite{Clementi,Clementi2008,Clementi2009,Clementi2010}, to model deterministic heterogeneity in tastes and product differentiation \cite{Rajaon,Rajaon2008} etc.

In the present contribution we reconsider critically the physical origin of the above mentioned statistical theory. Our main goal is to show that the function $\exp_{\kappa}(x)$,  and its inverse $\ln_{\kappa}(x)$, can be obtained within the one-particle relativistic dynamics, in a very simple and transparent way,  without invoking any extra principle or assumption, starting directly from the Lorentz transformations.

\sect{Lorentz Transformations and $\kappa$-exponential}

In the present section we will show that the function $\exp_{\kappa}(x)$ emerge within the special relativity as the relativistic generalization of the ordinary exponential of classical physics.
Special relativity is a theory founded on the Lorentz transformations, so
the question whether it is possible to obtain $\exp_{\kappa}(x)$
starting directly from the Lorentz transformations and how, arises
naturally.

Let us consider in the one-dimension frame $\cal S$ two identical
particles $A$ and $B$, of rest mass $m$. We suppose that the two
particles move with velocity $v_{\scriptscriptstyle A}$ and
$v_{\scriptscriptstyle B}$ respectively. The momenta of the two
particles are indicated with $p_{\scriptscriptstyle
A}=p\,(v{\scriptscriptstyle A})$ and $p_{\scriptscriptstyle
B}=p\,(v_{\scriptscriptstyle B})$, while their energies are
indicated with ${E}_{\scriptscriptstyle
A}={E}\,(v_{\scriptscriptstyle A})$ and ${E}_{\scriptscriptstyle
B}={E}\,(v_{\scriptscriptstyle B})$ respectively.

We consider now the same particles in the rest frame ${\cal S}'$ of
particle $B$. In this new frame the velocity, momentum and energy of
the particle $B$ are $v'_{\scriptscriptstyle B}=0$,
$p'_{\scriptscriptstyle B}=0$ and $E'_{\scriptscriptstyle B}=mc^2$
respectively. In ${\cal S}'$ the velocity of particle $A$ is given
by the formula $v'_{\scriptscriptstyle A}=(v_{\scriptscriptstyle
A}-v_{\scriptscriptstyle B})/(1-v_{\scriptscriptstyle A}
v_{\scriptscriptstyle B}/c^2)$ defining the relativistic velocity
composition law, which follows directly from the kinematic Lorentz
transformations. In the same frame ${\cal S}'$ the momentum and
energy of particle $A$ are given by the dynamic Lorentz
transformations
\begin{eqnarray}
&&p'_{\scriptscriptstyle A}=\gamma( v_{\scriptscriptstyle B})
p_{\scriptscriptstyle A}-c^{-2}v_{\scriptscriptstyle B}\gamma(
v_{\scriptscriptstyle B})E_{\scriptscriptstyle A} \ ,
\label{D1} \\
&&E'_{\scriptscriptstyle A}=\gamma( v_{\scriptscriptstyle B})
E_{\scriptscriptstyle A}-v_{\scriptscriptstyle B}\gamma(
v_{\scriptscriptstyle B})p_{\scriptscriptstyle A} \ , \label{D2}
\end{eqnarray}
$\gamma(v_{\scriptscriptstyle B})=(1-v_{\scriptscriptstyle
B}^2/c^2)^{-1/2}$ being the Lorentz factor.

We note that the expression of the momentum $p_{\scriptscriptstyle
B}=mv_{\scriptscriptstyle B}\gamma(v_{\scriptscriptstyle B})$ and of
the energy $E_{\scriptscriptstyle
B}=mc^2\gamma(v_{\scriptscriptstyle B})$ of the particle $B$ appear
explicitly in the latter transformations. After introducing in place
of the dimensional variables $(v,p,E)$ the dimensionless variables
$(u,q,{\cal E})$ through $v/u=p/mq=\sqrt{E/m{\cal E}}=|\kappa| c
=v_*<c$, the Lorentz transformations become
\begin{eqnarray}
&&q'_{\scriptscriptstyle A}=\kappa^2 q_{\scriptscriptstyle A} {\cal
E}_{\scriptscriptstyle B} - \kappa^2 q_{\scriptscriptstyle B} {\cal
E}_{\scriptscriptstyle A} \ ,
\label{D3} \\
&&{\cal E}'_{\scriptscriptstyle A}=\kappa^2 {\cal
E}_{\scriptscriptstyle A} {\cal E}_{\scriptscriptstyle B} -
q_{\scriptscriptstyle A} q_{\scriptscriptstyle B} \ , \label{D4}
\end{eqnarray}
while the classical limit $c\rightarrow \infty$ is replaced by the
limit $\kappa\rightarrow 0$. From the Lorentz invariance it follows
easily the energy-momentum dispersion relation $\kappa^4{\cal
E}^2-\kappa^2q^2=1$.

From its definition it follows that $\kappa$ is the Einstein $\beta$
factor related to the velocity $v_*$ and the condition  $v_*<c$
implies that $-1<\kappa <+1$. For a particle at rest it results
$E(0)=m\,c^2$ and then ${\cal E}(0)=1/\kappa^2$. Then $1/\kappa^2$
represents the dimensionless rest energy of the particle.
Alternatively $1/\kappa$ is the refractive index of a medium in
which the light speed is $v_*$.

It is remarkable  that the dynamic Lorentz transformations form a
system of two coupled linear equations, having the important feature
that the contributions of the two particles $A$ and $B$, appearing in
the right-hand side of the equations, are not factorized.
Spontaneously the question emerges at this point, whether new
variables exist, able to factorize the contribution of the two
particles $A$ and $B$ in the right-hand side of the dynamic Lorentz
transformations. For the kinematic  Lorentz transformations this
problem was first posed and solved by Minkowski, by introducing
light cone variables. After observing that both kinematic and
dynamic Lorentz transformations have the same structure, it is easy
to verify that the two dynamic light cone variables are given by
$\kappa^2 {\cal E} \pm \kappa\,q \geq 0$. Clearly any power of these
variables permits the factorization of the right-hand side of the Lorentz transformations and therefore we can write the more
general form of the dynamic Minkowski light cone variables as
follows
\begin{eqnarray}
\chi_{\pm}=\left(\kappa^2 {\cal E} \pm \kappa\,q\right )^{a} \ \
. \label{D5}
\end{eqnarray}
Lorentz transformations (\ref{D3}) and (\ref{D4}), after using the
variables $\chi_{\pm}$, assume the following
required form
\begin{eqnarray}
{\chi}'_{\pm A}={\chi}_{\mp B}\,{\chi}_{\pm A} \ \ ,
\label{D6}
\end{eqnarray}
while the condition expressing the Lorentz invariance of the system
simplifies to ${\chi}_{-}{\chi}_{+}=1$.

By taking into account the energy-momentum dispersion relation
$\kappa^4{\cal E}^2-\kappa^2q^2=1$, the light cone variables
${\chi}_{\pm}$ can be viewed as functions of the momentum $q$
i.e. ${\chi}_{\pm}={\chi}(\pm q)$. In order to completely
define the function ${\chi}(q)$ we need to fix the exponent $a$
appearing in (\ref{D5}). Only if we pose $a=1/\kappa$, the Lorentz
transformations can be reduced to the Galilei transformations, in
the classical $\kappa \rightarrow 0$ limit. Indeed, for
$a=1/\kappa$, it results $\lim_{\kappa \rightarrow
0}\,\chi(q)=\exp( q)$ and the Lorentz transformations (\ref{D6}) reduce
to
\begin{eqnarray}
\exp(\pm q'_{\scriptscriptstyle A})=\exp( \mp
q_{\scriptscriptstyle B}) \exp(\pm q_{\scriptscriptstyle A})
\ \ , \label{D7}
\end{eqnarray}
so that the Galilei transformation for the momenta,
$q'_{\scriptscriptstyle A} = q_{\scriptscriptstyle A}
-\,q_{\scriptscriptstyle B}$ follows immediately.

After posing $a=1/\kappa$, and taking into account the dispersion relation ${\cal E}=\sqrt{1+\kappa^2 q^2}/\kappa^2$, Eq.(\ref{D5}) univocally
defines the dynamic Minkowski light cone variables $\chi(q)$,
which interestingly, coincides with the $\kappa$-exponential
given by (\ref{A1}), i.e.
\begin{eqnarray}
{\chi}(q)=\exp_{\kappa}\!\left( q\right)
\ . \label{D8}
\end{eqnarray}

Finally the Lorentz transformations (\ref{D6}) assume the
required, factorized simple form
\begin{eqnarray}
\exp_{\kappa}(\pm q'_{\scriptscriptstyle A})=\exp_{\kappa}( \mp
q_{\scriptscriptstyle B}) \exp_{\kappa}(\pm q_{\scriptscriptstyle A})
\ \ . \label{D9}
\end{eqnarray}

It is remarkable that the function $\exp_{\kappa}(x)$, defining the particle distribution function in statistical physics, emerges also in one-particle special relativity as the light cone variable, able to factorize the Lorentz transformations, and represents the relativistic generalization of the ordinary exponential.

The Galilei relativity principle, holding both in classical physics
and in special relativity, imposes the equivalence of all the
inertial frames. According to this principle, the inverse Lorentz
transformations must have the same structure of the direct
transformations (\ref{D9}) except for the substitutions
$q'_{\scriptscriptstyle A} \leftrightarrow q_{\scriptscriptstyle A}$
and $q_{\scriptscriptstyle B} \rightarrow - q_{\scriptscriptstyle
B}$. This requirement implies for the generalized exponential the
property
\begin{eqnarray}
\exp_{\kappa}(x)\exp_{\kappa}(-x)= 1 \ \ ,
\label{D10}
\end{eqnarray}
which in terms of the generalized logarithm becomes
\begin{eqnarray}
\ln_{\kappa}(1/x)= - \ln_{\kappa}(x) \ \ . \label{D11}
\end{eqnarray}
The latter relationships imposed directly by the Galilei relativity principle, represents two important  properties of the functions $\exp_{\kappa}(x)$ and $\ln_{\kappa}(x)$ respectively.

\sect{Lorentz Transformations and $\kappa$-logarithm}

In the present section we will show that the function $\ln_{\kappa}(x)$, can be obtained by employing the Lorentz transformations only.

Let us consider the Lorentz transformations in the case where $q_{_B}=q$, $q_{_A}=q+dq$ and $q'_{_A}=d_{\kappa}q$. Eq. (\ref{D9}) becomes
\begin{eqnarray}
\exp_{\kappa}\!\left(d_{\kappa}q \right)\,\,
=\exp_{\kappa}\!\left( - q \right)
\exp_{\kappa}\!\left( q + d q \right)  \ \ .
\label{E3}
\end{eqnarray}
By performing a Taylor expansion, up to the first order, in the latter equation and after taking into account (\ref{D10}) and the normalization condition $\exp'_{\kappa}(0)= 1$ we obtain
\begin{eqnarray}
\frac{d\,\exp_{\kappa}\!\left( q \right)}{d_{\kappa}q}=
\exp_{\kappa}\!\left( q\right)  \ \ .
\label{E4}
\end{eqnarray}

Preliminary, we observe that the composition law of the relativistic momenta can be established starting from the first equation of the Lorentz transformations (\ref{D3}), after elimination of the energy by taking into account the dispersion relation ${\cal E}=\sqrt{1+\kappa^2q^2}/\kappa^2$, obtaining
\begin{eqnarray}
\displaystyle{q'_{_A}=q_{_A}\sqrt{1+\kappa^2q_{_B}^2}-q_{_B}\sqrt{1+\kappa^2q_{_A}^2}}
 \ \ . \
  \label{E5}
\end{eqnarray}

The expression of the $\kappa$-differential $d_{\kappa}q$, can be obtained starting directly from the latter composition law, by posing  $q_{_B}=q$, $q_{_A}=q+dq$ and $q'_{_A}=d_{\kappa}q$ and after performing a Taylor expansion i.e.
\begin{eqnarray}
d_{\kappa}q=\frac{dq}{\gamma_{\kappa}(q)}
 \ \ , \
  \label{E6}
\end{eqnarray}
$\gamma_{\kappa}(q)=\sqrt{1+\kappa^2q^2}$ being the Lorentz factor. After taking into account (\ref{E6}), Eq. (\ref{E4}) becomes
\begin{eqnarray}
\gamma_{\kappa}(q)\,\frac{d\,\exp_{\kappa}\! q }{dq}=
\exp_{\kappa}\! q  \ \ .
\label{E7}
\end{eqnarray}
The differential equation for the function $\ln_{\kappa}(f)$ can be obtained by performing in Eq. (\ref{E7}) the change of variable $q=\ln_{\kappa}(f)$,
\begin{eqnarray}
\frac{d\,\ln_{\kappa}\! f }{df} =\frac{\gamma_{\kappa}\big(\ln_{\kappa}f \big)}{f}
  \ \ .
\label{E8}
\end{eqnarray}

At this point after posing $\exp_{\kappa}(q_{_A})=f$, $\exp_{\kappa}(q'_{_A})=f'$, and $\exp_{\kappa}(-q_{_B})=\epsilon$, the Lorentz transformations in the factorized form (\ref{D9}), become $f'=\epsilon f$. On the other hand the first of the Lorentz transformations (\ref{D3}) yields
\begin{eqnarray}
\ln_{\kappa}(\epsilon f)= \gamma_{\kappa}\big(\ln_{\kappa}\epsilon\big)\,\ln_{\kappa}f
+ \gamma_{\kappa}\big(\ln_{\kappa}f\big)\,\ln_{\kappa}\epsilon
\ , \ \
\label{E9}
\end{eqnarray}
so that we obtain that $\gamma_{\kappa}\big(\ln_{\kappa}f \big)$ is given by
\begin{eqnarray}
 \gamma_{\kappa}\big(\ln_{\kappa}f\big)\,=
 \frac{\ln_{\kappa}(\epsilon f)}{\ln_{\kappa}\epsilon}- \frac{\gamma_{\kappa}\big(\ln_{\kappa}\epsilon\big)}{\ln_{\kappa}\epsilon}
\,\ln_{\kappa}f
\ . \ \
\label{E10}
\end{eqnarray}
Finally, by accounting (\ref{E10}) in Eq. (\ref{E8}) we obtain

\newpage

\begin{equation}
\frac{d}{df}\,f\,\ln_{\kappa}\! f   = \frac{\ln_{\kappa}(\epsilon f)}{\ln_{\kappa}\epsilon} \, +
\left [1 \!-\! \frac{\gamma_{\kappa}\big(\ln_{\kappa}\epsilon\big)}{\ln_{\kappa}\epsilon} \right ]\ln_{\kappa}f
  \ . \ \ \ \ \ \
\label{E11}
\end{equation}

After choosing $v_*=-v_{_B}$, which implies $u_{_B}=-1$, it results $q_{_B}=-1/\sqrt{1-\kappa^2}$ and therefore $\ln_{\kappa}\epsilon=1/\sqrt{1-\kappa^2}$. The latter relationship simplifies  Eq. (\ref{E11}) as follows
 \begin{equation}
\frac{d}{df}\,f\,\ln_{\kappa}\! f  = \frac{\ln_{\kappa}(\epsilon f)}{\ln_{\kappa}\epsilon} \,
  \ . \ \ \ \ \ \
\label{E12}
\end{equation}
Eq. (\ref{E12}), already known in the literature, has been introduced heuristically in \cite{PRE02}. Its solution, with the condition (\ref{D11}), is unique and defines the generalized logarithm $\ln_{\kappa}(f)$, as given in Eq. (\ref{A2}).
Here, Eq. (\ref{E12}), has been obtained by employing a physical mechanism based, exclusively, on the Lorentz transformations and without invoking any extra principle or assumption.

\sect{Conclusions}

As conclusions we recall the main results obtained in the
present Letter:

(i) We have shown that the Lorentz transformations are able to furnish a very transparent theoretical support to the function $\exp_{\kappa}(x)$. This function results to be a light cone variable and therefore factorizes the energy-momentum Lorentz transformations.

(ii) We have shown that the mathematical property $\exp_{\kappa}(x)\exp_{\kappa}(-x)=1$ is enforced exclusively by the Galilei relativity principle.

(iii) Eq. (\ref{E12}), introduced heuristically in the literature after 2002,  whose solution defines the function $\ln_{\kappa}(x)$, has been obtained here starting directly from the Lorentz transformations.

\newpage


\begin{thebibliography}{99}





\bibitem{PRE02}
G. Kaniadakis,
Phys. Rev. E {\bf 66},  056125 (2002).
\bibitem{PRE05}
G. Kaniadakis,
Phys. Rev. E {\bf 72}, 036108 (2005).
\bibitem{EPL10}
G. Kaniadakis,
Europhys. Lett. {\bf 92}, 35002 (2010).
\bibitem{EPJB2009}
G. Kaniadakis,
Eur. Phys. J. B {\bf 70}, 3 (2009).
\bibitem{PhA01}
G. Kaniadakis,
Physica A {\bf 296}, 405 (2001).
\bibitem{PLA01}
G. Kaniadakis,
Phys. Lett. A {\bf 288}, 283 (2001).
\bibitem{EPJA2009}
G. Kaniadakis,
Eur. Phys. J. A {\bf 40}, 275 (2009).


\bibitem{Silva06A}
R. Silva,
Eur. Phys. J. B {\bf 54}, 499 (2006).
\bibitem{Silva06B}
R. Silva,
Phys. Lett. A {\bf 352} 17 (2006).

\bibitem{Wada1}
T. Wada,
Physica A {\bf 340}, 126 (2004).
\bibitem{Wada2}
T. Wada,
Contin. Mechan. and Thermod. {\bf 16}, 263 (2004).

\bibitem{KSPA04}
G. Kaniadakis, A.M. Scarfone,
Physica A {\bf 340}, 102 (2004).
\bibitem{AKSJPA04}
S. Abe, G. Kaniadakis and A.M. Scarfone,
J. Phys. A: Math. Gen. {\bf 37}, 10513 (2004).
\bibitem{Naudts1}
J. Naudts,
Physica A {\bf 316}, 323 (2002).
\bibitem{Naudts2}
J. Naudts,
Rev. Math. Phys. {\bf 16}, 809 (2004).

\bibitem{ScarfoneWada}
A.M. Scarfone, T. Wada,
Progress of Theor. Phys. Suppl. {\bf 162} 45 (2006).

\bibitem{Yamano}
T. Yamano,
Phys. Lett. A {\bf 308}, 364 (2003).

\bibitem{Pistone}
G. Pistone,
Eur. Phys. J. B {\bf 70}, 29 (2009).

\bibitem{OikPOMP10}
T. Oikonomou, G. Baris Bagci,
Rep. Math. Phys. {\bf 66}, 137 (2010).


\bibitem{GuoRelativistic}
Guo Lina, Du Jiulin, and Liu Zhipeng,
Phys. Lett. A {\bf 367}, 431 (2007).
\bibitem{GuoClassic}
Guo Lina and Du Jiulin,
Phys. Lett. A {\bf 362}, 368 (2007).


\bibitem{Lapenta}
G. Lapenta, S. Markidis, A. Marocchino, and G. Kaniadakis,
Astrophysical Journal {\bf 666}, 949 (2007).
\bibitem{Lapenta2009}
G. Lapenta, S. Markidis, G. Kaniadakis,
J. of Stat. Mech., P02024 (2009).

\bibitem{WadaScarfone2009}
T. Wada, A.M. Scarfone,
Eur. Phys. J. B {\bf 70}, 29 (2009).
\bibitem{Wada2010}
T. Wada,
Eur. Phys. J. B {\bf 73}, 287 (2009).


\bibitem{Rossani}
A. Rossani and A.M. Scarfone,
J. Phys. A {\bf 37}, 4955 (2004).

\bibitem{GuoDuoTgradient}
Guo L.N., Du J.L.,
Physica A {\bf 389}, 47-51 (2010).


\bibitem{Silva2008}
J.M. Silva, R. Silva, J.A.S. Lima,
Phys. Lett. A {\bf 372}, 5754 (2008).

\bibitem{Carvalho}
J. C. Carvalho, R. Silva, J.D. do Nascimento jr., and J. R. De Medeiros,
Europhys. Lett.  {\bf 84}, 59001 (2008).
\bibitem{Carvalho2}
J. C. Carvalho, J.D. do Nascimento jr., R. Silva, and J. R. De Medeiros,
Astrophysical Journal Letters 696, L48 (2009).
\bibitem{CarvalhoEPL10}
J. C. Carvalho, R. Silva, J. D. do Nascimento jr., B. B. Soares and J. R. De Medeiros,
Europhys. Lett. {\bf 91}, 69002 (2010).


\bibitem{Tewel}
A.M. Teweldeberhan, H.G. Miller, and R. Tegen,
Int. J. Mod. Phys. E {\bf 12}, 669 (2003).

\bibitem{Pereira}
F.I.M. Pereira, R. Silva, J.S. Alcaniz,
Nucl. Phys. A {\bf 828}, 136 (2009).


\bibitem{Fracture}
M. Cravero, G. Iabichino, G. Kaniadakis, E. Miraldi, A.M. Scarfone,
Physica A {\bf 340}, 410 (2004).


\bibitem{Corradu}
M. Coraddu, M. Lissia, R. Tonelli,
Physica A {\bf 365}, 252
(2006).
\bibitem{Tonelli}
R. Tonelli, G Mezzorani, F. Meloni, M. Lissia, M. Coraddu.
Prog. Theor. Phys. {\bf 115},  23  (2006).
\bibitem{Celikoglu}
A. Celikoglu A, U. Tirnakli,
Physica A {\bf 372}, 238 (2006).

\bibitem{Olemskoi}
A.I. Olemskoi, V.O. Kharchenko, V.N. Borisyuk,
Physica A {\bf 387}, 1895 (2008).
\bibitem{OlemskoiEPJB10}
A.I.Olemskoi, S.S. Borysov, I.A. Shuda,
Eur. Phys. J. B {\bf 77}, 219 (2010).


\bibitem{AbulMagd}
A.Y. Abul-Magd,
Phys. Lett. A {\bf 361}, 450 (2007).
\bibitem{AbulMagd2009}
A.Y. Abul-Magd,
Eur. Phys. J. B {\bf 70}, 39 (2009).



\bibitem{WadaSuyari06}
T. Wada, H. Suyari,
Phys. Lett. A {\bf 348}, 89 (2006).

\bibitem{Topsoe}
F. Topsoe,
Physica A {\bf 340} 11 (2004).

\bibitem{WadaSuyari07}
T. Wada, H. Suyari,
Phys. Lett. A {\bf 368}, 199 (2007).

\bibitem{Clementi}
F. Clementi, M. Gallegati, G. Kaniadakis,
Eur. Phys. J. B {\bf 57}, 187  (2007).
\bibitem{Clementi2008}
F. Clementi, T. Di Matteo, M. Gallegati, G. Kaniadakis,
Physica A {\bf 387}, 3201 (2008).
\bibitem{Clementi2009}
F. Clementi, M. Gallegati, G. Kaniadakis,
J. of Stat. Mech., P02037 (2009).
\bibitem{Clementi2010}
F. Clementi, M. Gallegati, G. Kaniadakis,
Emp. Econ. {\bf 39}, 559 (2010).



\bibitem{Rajaon}
D. Rajaonarison, D. Bolduc, and H. Jayet,
Econ. Lett. {\bf 86}, 13  (2005).
\bibitem{Rajaon2008}
D. Rajaonarison,
Econ. Lett. {\bf 100}, 396 (2008).





\end{thebibliography}
\end{document}